\documentclass[showpacs,twocolumn,amsmath,amssymb,prb]{revtex4}
\usepackage{graphicx}
\usepackage{epsfig,color}

\begin{document}
\title{Josephson effect and Andreev reflection in Ba$_{1-x}$Na$_{x}$Fe$_{2}$As$_{2}$ ($x$=0.25 and 0.35) point contacts}

\author{V. V. Fisun$^1$}
\author{O. P. Balkashin$^1$}
\author{O. E. Kvitnitskaya$^1$}
\author{I. A. Korovkin$^1$}
\author{N.V. Gamayunova$^1$}
\author{S. Aswartham$^2$}
\author{S. Wurmehl$^2$}
\author{Yu. G. Naidyuk$^1$}


\affiliation{$^1$B.\,Verkin Institute for Low Temperature Physics and Engineering, National Academy of Sciences of Ukraine,
47 Lenin Ave., 61103, Kharkiv, Ukraine}

\affiliation{$^2$Leibniz-Institut f\"{u}r Festk\"{o}rper- und Werkstoffforschung Dresden e.V., Postfach 270116, D-01171 Dresden, Germany}

\begin{abstract}

$I(V)$ characteristics and their first derivatives of ScS and ScN-type (S--superconductor,
c--constriction, N--normal metal) point-contacts (PCs) based on Ba$_{1-x}$Na$_{x}$Fe$_{2}$As$_{2}$
($x$=0.25 and 0.35) were studied. ScS-type PCs with S=Nb,Ta and Pb show
Josephson-like resistively shunted $I(V)$ curves with microwave induced Shapiro steps which satisfy
relation 2$eV=\hbar\omega$. The $I_cR_N$ product ($I_c$--critical current, $R_N$-- normal state PC resistance)
in these PCs is found to be up to 1.2\,mV.
All this data with the observed dependence of the $I_c$ on the microwave power of ScS PCs with Pb
counterelectrode indicates the presence of the singlet s-wave type pairing in Ba$_{1-x}$Na$_{x}$Fe$_{2}$As$_{2}$.
From the  $dV/dI(V)$ curves of ScN-type PCs demonstrating Andreev-reflection like features,
the superconducting gap $\Delta $ ratio 2$\Delta/$k$_B$T$_c$ = 3.6$\pm$1 for the compound with $x$=0.35
was evaluated. Analysis of these  $dV/dI(V)$ at high biases $V$, that is well above $\Delta $,
testifies transition to the thermal regime in PCs with a voltage increase.


\end{abstract}


\maketitle

\section{INTRODUCTION}

The discovery of superconductivity in the iron-based materials provokes an enormous interest,
since a high $T_c$  was obtained in compounds with ferromagnetic metals, what raised the question
about nature of superconducting (SC) pairing mechanism and symmetry of SC wave function.
Ferriferous doped superconductors of 122-type structure based on the parent compound BaFe$_2$As$_2$
are the most investigated pnictide systems nowadays. Hole doping by alkali
metals leads to the suppression of the spin-density wave antiferromagnetic state and
the appearance of superconductivity with transition temperature $T_c$ up to 38\,K \cite{Rotter2008,Sasmal}.
Superconductivity is formed here in a multi-band system with multiple Fermi surfaces of different
(electron and hole) nature, which is very different from the single band situation in high-$T_c$  cuprates.
Since their discovery, quite a lot of research has been done, in which the electron spectrum and the Fermi surface of new
superconductors were studied using angle-resolved photoemission (ARPES) \cite{Kordyuk}.
This method has proved its effectiveness in the physics of high-$T_c$ cuprates.
In fact, for iron-based superconductors  ARPES-study provides valuable information that helps
clarify the peculiarities of their electronic spectrum, as well as of the Fermi surface,
and the quantities and characteristics of SC gap(s).
Investigation of the Josephson effect \cite{Seidel} and point-contact Andreev-reflection (PCAR)
spectroscopy \cite{Daghero2011} plays significant role in the understanding of the nature
of the SC ground state in these material as well.

The Josephson effect \cite{KY,Barone} was investigated in pressure-type contacts with
the Pb counterelectrode in \cite{Zhang09}. Stable Josephson coupling in contacts
Ba$_{1-x}$Na$_{x}$Fe$_{2}$As$_{2}$--Pb along the c-axis was found.
It was established that a Josephson current flows mainly through the active small contacts
which exhibit virtually no superconductivity at low clamping. Just as the pressure increases,
the contact area increases and superconductivity appears.
Observation of the Josephson effect in the c-axis geometry in this iron pnictides excludes
pure p- or d-wave pairing in these materials and the obtained results support the existence of s-wave
pairing \cite{Zhang09a}.
Josephson junctions fabricated in epitaxial films utilizing bicrystal grain boundary \cite{Katase}
and oxidized titanium layers as barriers \cite{Doring} have been studied as well.
Both studies report also some difficulty to get junctions with Josephson behaviour
and the product $I_cR_N$ is up to several orders of magnitude less than that expected theoretically,
what is a serious obstacle for an application.

On the other hand, PCAR spectroscopy is widely used to study SC order parameter(s) or SC gap(s)
in these compounds \cite{Daghero2011,Szabo,XinLu}. In our previous paper \cite{NaidyukBanafeas}
we utilized the PCAR spectroscopy to study the SC gap in the sample with $x$=0.25. It was shown
that pronounced peculiarities in the PC spectra at high biases above the SC gap
is governed by high specific resistivity and thermopower of the bulk material being
the features of the thermal regime of the current flow in PCs.
Whereas emerging of AR structures at small biases gave possibility for the spectroscopy
of the SC gap which showed unequivocally only one gap with the preferred value 2$\Delta/$k$_{B}T_{c}\approx$6.
In this study we continue PCAR investigation of the Ba$_{1-x}$Na$_{x}$Fe$_{2}$As$_{2}$ family,
focusing on the compound with $x$=0.35, in order to have a look on the gap structure
in this compound and compare with the results obtained in \cite{NaidyukBanafeas} for $x$=0.25.

\section{EXPERIMENTAL DETAILS}

Single crystals of Ba$_{1-x}$Na$_{x}$Fe$_{2}$As$_{2}$ were grown using a self-flux high temperature solution growth technique \cite{Aswartham}.
For the samples with $x$ = 0.25 and 0.35 the SC transition in resistivity starts around 10 and 34\,K, correspondingly.

The PCs were established \textit{in situ} by touching a thin metallic wire (\O $\approx$0.2--0.3\,mm) to the cleaved
(at room temperature) surface (an edge) of the plate-like (flake) sample. Thus, we measured heterocontacts
between simple metal and the title compound preferably in the $ab$-plane. The differential resistance $dV/dI(V)\equiv R(V)$
of $I(V)$ characteristic of PC was recorded by sweeping the \textit{dc} current $I$ on which a small \textit{ac}
current $i$ was superimposed using standard lock-in technique. The measurements were performed mainly at
the temperature 4.2\,K (up to 35\,K in some cases) and under microwave irradiation with frequency 9.57\,GHz
in the case of ScS contacts.

The critical current $I_c$ (Josephson effect) was registered on $I(V)$ curves of ScS contacts
only after electrical breakdown of mechanically established PC from resistance of a few
tenth Ohm till much lower resistance, i.e. much less than 1$\Omega$. For PCs with Pb counterelectrode
we were able to produce Josephson junction with a clear critical current or zero resistance
at $V$=0, while in the case of the Ta and Nb counterelectrodes an additional
resistance appeared, so that $R(V=0)\neq0$.

The characteristic voltage $V = I_cR_N$ ($R_N$ is normal state resistance of PC)
that determines the cutoff frequency of the Josephson current \cite{Likharev},
was up to  $V$=1.2\,meV for the PCs with Pb counterelectrode. Induced Shapiro
steps were observed under microwave irradiation both on $I(V)$ curves and more clearly on their
first derivatives with the distance between them $V=\hbar\omega/2e$.

For some ScS contacts with the Pb counterelectrode, as well as for all contacts with the Ta counterelectrode,
the critical current was very small and the manifestation of the $ac$ Josephson effect (Shapiro steps) was
observed only on the first derivatives of the $I(V)$. Equivalent circuit for PCs in this case, can be
represented as a series connection of the resistance with a Josephson junction. In the case of Nb counterelectrode
a more complex circuit was implemented with serial and parallel additional resistances as it will be shown further.

\section{RESULTS AND DISCUSSION}
\subsection{Search for Josephson effect in ScS contacts}
Figure 1 shows a series of the $I(V)$ curves and first derivative of a ScS PC Ba$_{1-x}$Na$_{x}$Fe$_{2}$As$_{2}$
with Pb counterelectrode at different microwave power levels. It is noticeable that under irradiation
a stepped structure occurs in the $I(V)$. This is,  so-called, Shapiro steps, or manifestation of the $ac$
Josephson effect \cite{KY,Barone}.
\begin{figure}[t]
\includegraphics[width=9cm,angle=0]{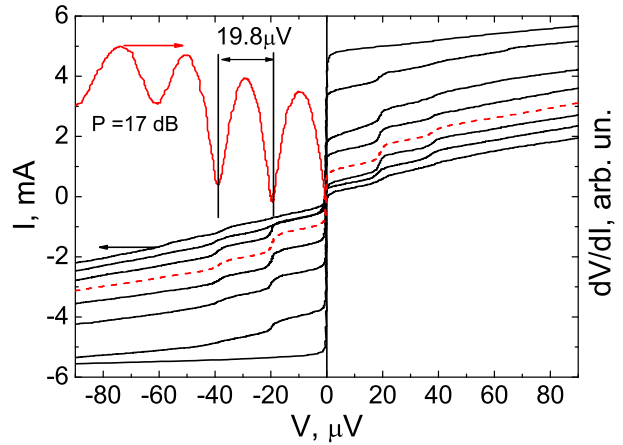}
\caption{ (Color online)
$I(V)$ curves of contact Ba$_{0.65}$Na$_{0.35}$Fe$_{2}$As$_{2}$ -- Pb
at increasing of microwave power from zero (top curve) to the maximal (bottom curve). For the
dotted (red) $I(V)$ curve its first derivative $dV/dI(V)$ is shown for negative bias. $T$ = 4.2\,K, $f$ = 9.57\,GHz.}
\label{fig1}
\end{figure}

For some contacts with the Pb counterelectrode (Fig.\,2) there was a significant slope
of $I(V)$ at $V$=0 and the critical current was small. This leads to the fact that
the distance between the steps becomes larger than the expected one. The $I(V)$ slope at $V$=0 indicates
the inclusion of an additional resistance $R_{ad}$ in series with the Josephson junction
(see the equivalent circuit in Fig.\,2). Accounting of this resistance ($R_{ad}$=28\,m$\Omega$)
leads to the elimination of $I(V)$ inclination at $V$=0 (Fig.\,2 inset),
while appropriate distance between steps Shapiro becomes $V=\hbar\omega/2e$
as it should be in the case of Josephson coupling.
It should be noted that the induced current step with the largest number was observed at $V \sim 1$\,mV.
This value coincides with the characteristic voltage estimated from the product of $I_cR_N$.
\begin{figure}[t]
\includegraphics[width=9cm,angle=0]{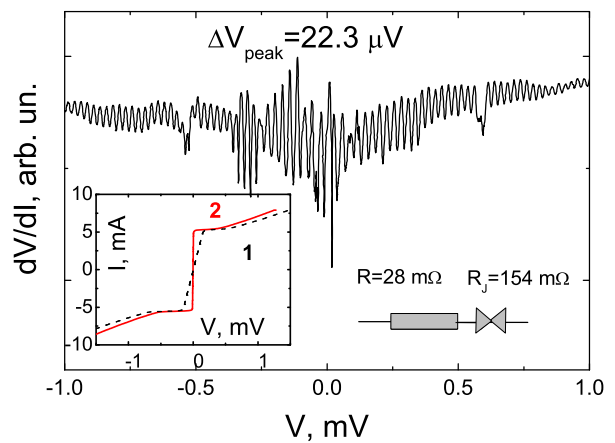}
\caption{ (Color online)
First derivative of $I(V)$ curve of  Ba$_{0.25}$Na$_{0.25}$Fe$_{2}$As$_{2}$ -- Pb contact.
Inset: experimental $I(V)$ (dotted curve 1),  solid curve 2 is  $I(V)$ after subtracting
of additional resistance.}
\label{fig2}
\end{figure}

The position of current steps on the voltage axis is defined by the Josephson relation  \cite{KY,Barone}:
\begin{equation}
\label{eq1}
V_{n} =nhf/2e,n=0,1,2,3,...
\end{equation}
with a superconducting pair charge equal to 2e.
It is known  \cite{KY,Barone} that the height of the current steps induced  by the external field
oscillates with increase of the irradiation power. The amplitude of the current steps
is given by the Bessel function of the corresponding order  \cite{KY,Barone}:
\begin{equation}
\label{eq2}
I_{n} =I_{0} \left| {J_{n} (e\nu/hf)\left| \right.} \right. ,
\end{equation}
where  $\nu \sim \sqrt P $ ($P$ is microwave power) is $ac$ voltage induced in a contact by microwave field.
Fig.\,3 shows  dependence of the critical current $I_c$ and the first two steps of currents $I_1$ and $I_2$
on the microwave power. The resulting curves are in good agreement with
the initial portions of the corresponding Bessel functions \cite{Barone}.
Since the investigated Josephson junctions have low resistance (R$_{N}<0.2\Omega$) then we do not have
enough microwave power to trace the behavior of the critical current $I_c$ and the first two steps
of the currents $I_1$ and $I_2$ to a greater extent.

\begin{figure}[t]
\includegraphics[width=6cm,angle=0]{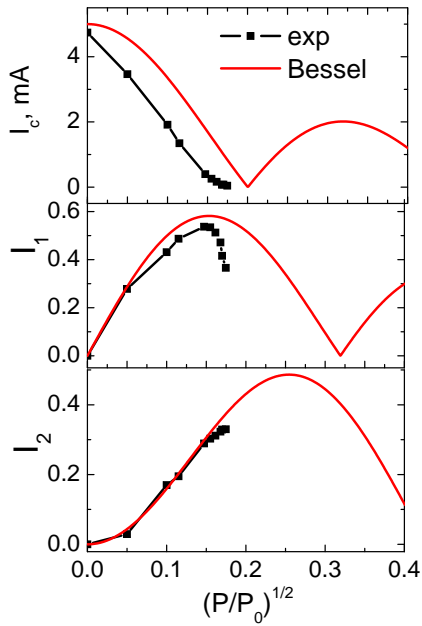}
\caption{ (Color online)
Dependence of the critical current $I_c$ and the amplitude of the first microwave induced steps
of currents $I_1$, $I_2$ on microwave power $P$.}
\label{fig3}
\end{figure}

In the case of the Ta counterelectrode, electrical breakdown is also used to create a Josephson junction.
However, in this case, there was always a significant slope of  $I(V)$  at $V$=0
and thus the critical current was negligible. This leads to the observation that Shapiro steps become
visible only on the first derivative $dV/dI(V)$ as dips and the distance between them was greater
than the expected one (Fig.\,3). Thus, for the ScS contacts with the Ta counterelectrode the equivalent circuit
includes a series resistance, as well as for some PCs with Pb. Accounting for this resistance
$R_{ad}=21.7$m$\Omega$ leads to the elimination of $I(V)$ inclination at $V$=0
and the distance between Shapiro steps becomes correct, i.e. $V=\hbar\omega/2e$.

\begin{figure}[t]
\includegraphics[width=8cm,angle=0]{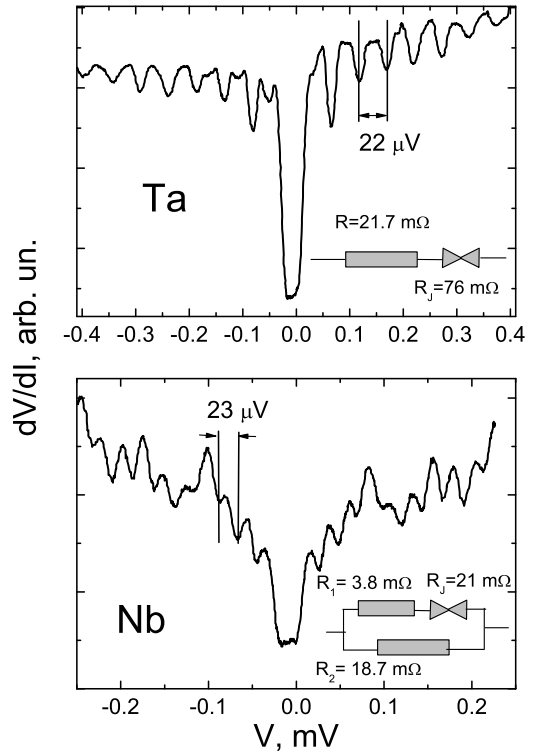}
\caption{ Upper panel:
The first derivative of $I(V)$ for a ScS contact Ba$_{0.65}$Na$_{0.35}$Fe$_{2}$As$_{2}$--Ta at $T$=4.2\,K,
$f$=9.57\,GHz. Shapiro steps appear as dips spaced by 22$\mu $V.
The inset shows the equivalent circuit of a Josephson junction with serial resistance accounting for additional resistance.
Bottom panel:
The first derivative of $I(V)$ for a ScS contact Ba$_{0.65}$Na$_{0.35}$Fe$_{2}$As$_{2}$--Nb at $T$=4.2\,K, $f$=9.57\,GHz.
Shapiro steps appear as dips spaced by 22$\mu $V.
The insets shows the equivalent circuit of a Josephson junction with the additional serial and parallel resistances.}
\label{fig4}
\end{figure}

Josephson junctions with a counter electrode made of Nb, as well as in the case of Ta and Pb, had also a slope
of  $I(V)$ at $V$=0 and a larger distance between Shapiro steps. The equivalent circuit here is more complicated,
since the incorporation of additional resistance in the circuit with series connection leads to overestimation
of distance between the steps. The equivalent circuit of a ScS contact in this case, following \cite{Balk90},
can be represented as a parallel circuit, in one arm, which includes a Josephson junction in series
with the resistance $R_1$, and the other - an additional parallel resistance $R_2$. It is assumed
that the resistive elements have a linear $I(V)$ and the power of the incident radiation is sufficiently
low to prevent bolometric effect, i.e. the contact heating by irradiation is negligible.


Applying the methodology outlined in \cite{Balk90} to our ScS contacts with Ta and Nb,
we show the final electrical circuits in  Fig.\,4 (insets).

The absence of the critical current (nonzero resistance at $V$=0) on $I(V)$ in the case of the Ta and Nb counterelectrodes
makes it impossible to trace its behavior as a function of the microwave power. Difficulties associated with
the creation of Josephson junctions with high critical parameters in the case of Ta and Nb counterelectrodes,
can be connected with sensitivity of the SC parameters of pnictides to stoichiometry of the surface.
It may be associated with degradation of the surface structure of Ba$_{1-x}$Na$_{x}$Fe$_{2}$As$_{2}$ in the contact.
Namely, Pb is a softer counterelectrode compared with Ta and Nb, what reduces the surface damage at the PC creating

Thus, the detection of the ac Josephson effect in point contacts between  Ba$_{1-x}$Na$_{x}$Fe$_{2}$As$_{2}$
crystals and Pb, Ta and Nb  counterelectrode with the distance between  Shapiro steps $V=\hbar\omega/2e$,
as well as the observed dependence of the critical current on the microwave power with Pb counterelectrode
is in line with the presence of the singlet s-wave pairing in this new superconductor.

\subsection{Andreev reflection spectroscopy using ScN contacts}

We have measured and analyzed a few tens of \textit{dV/dI(V)} dependences of
Ba$_{0.75}$Na$_{0.35}$Fe$_2$As$_2$ -- Ag (or Cu) ScN-type PCs. As in the case of the samples with
$x$=0.25 \cite{NaidyukBanafeas}, the $dV/dI(V)$ spectra do not show any principal difference while being
measured by attaching the needle to the cleaved surface or to an edge of the samples.
While measuring different PCs below transition temperature T$_c$ the various shapes of \textit{dV/dI} were observed.
They are shown in Fig.\,\ref{fig1} for several PCs. Let us consider peculiarities of $dV/dI(V)$
in Fig.\,\ref{fig1}. Some of them are supposed to be due to Andreev reflection effect.
These are the double \textit{dV/dI} minima at energies roughly corresponding to the SC energy gap.
These features are visible in two upper \textit{dV/dI} characteristics shown in Fig.\,\ref{fig1}.
The position of the \textit{dV/dI} minima appear to be in the range between 5 and 15\,meV
for different PCs. Whereas two lower PCs in Fig.\,\ref{fig1} are presumably in
the thermal limit of the current flow and their \textit{dV/dI} shape is caused by the bulk resistivity
and thermal effects as mentioned in \cite{NaidyukBanafeas} for compound with $x$=0.25.
The middle curve shows the shallow AR-like minima around $V$=0 followed by spikes, while its further behavior
is similar to that of two bottom curves at large bias. That is this spectrum looks like transitional one
between the upper and the lower spectra. The much larger change of the differential resistance for two lower
\textit{dV/dI} also confirms the realization of the thermal regime for these PCs.
We should also mention that unlike the system with $x=0.25$ investigated earlier \cite{NaidyukBanafeas},
where the local $T_{c}$ in PCs had a large dispersion, i.e. from 10\,K up to 20\,K and higher, the actual samples
showed a rather stable local $T_c$ being around 30\,K. Also the \textit{dV/dI} curves of the samples
with $x=0.35$ did not display a pronounced Kondo-like shape above $T_c$ as some of those of the specimen with $x=0.25$.
All this points to a better homogeneity and quality of the investigated samples with $x=0.35$ at least
on the surface.

\begin{figure}[t]
\includegraphics[width=8cm,angle=0]{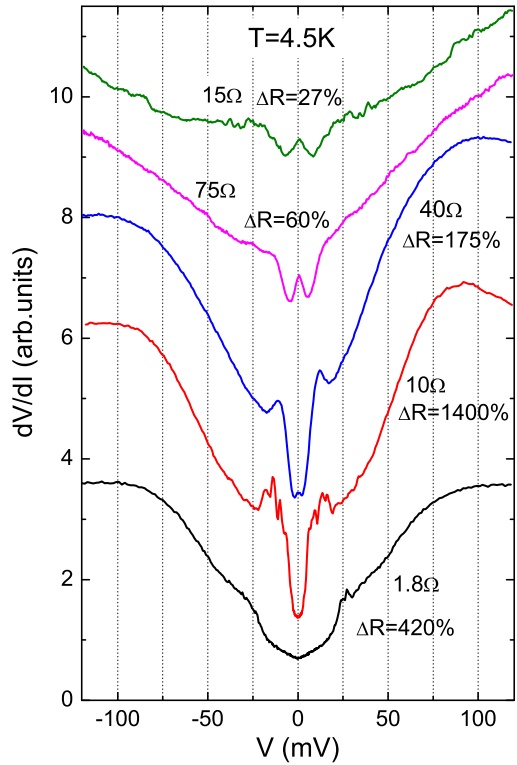}
\caption{(Color online) \textit{dV/dI(V)} curves displaying the SC features (deep minima)  around zero-bias
for the PCs with different resistance $R$ ($R$ is measured  just above the SC minimum at $V > \Delta/e)$).
The change $\Delta R=(R(V=100\,$mV$)-R(V=0))/R(V=0)$ of the PC resistance is shown in per cents for each PC.
The curves are shifted for the clarity.}
\label{fig6}
\end{figure}

The necessary condition for receiving of the spectroscopic information from the PC data (among
them is the SC gap) is the fulfilment of the spectroscopic (nonthermal) regime of the current flow
through PC \cite{PCSbook}. In this case the size (diameter  $d$) of PCs should be less than inelastic
mean free path $l_i$  of electrons and an additional requirement $d<\xi$ for PCs ($\xi$ is the coherence length)
is desirable to prevent the variation of the order parameter (SC gap) in the PC core.
As we mentioned in \cite{NaidyukBanafeas}, the elastic electronic mean-free path in Ba$_{1-x}$Na$_{x}$Fe$_{2}$As$_{2}$
is small (to be about  10\,nm)
and the coherence length $\xi$ amounts to only 2\,nm in the isostructural system Ba$_{1-x}$K$_{x}$Fe$_{2}$As$_{2}$ \cite{Wray}.
Evaluation of the typical PC diameter for investigated PCs, as made in \cite{NaidyukBanafeas},
gives for the PC size values between 2 and 200\,nm for the PC resistances from 1 to 110\,$\Omega$.
Thus, only for very high-ohmic PCs the ballistic (spectroscopic) regime can be realized. Therefore, we suppose the possibility of
the diffusive regime for low-ohmic PCs at small voltage biases. Another criterion of the spectral
regime is the observation of AR-like features in the \textit{dV/dI(V)} curves in the SC state.

One of the main tasks at the interpretation PC data in systems with very short electronic lengths is the separation
of the "thermal" features from the spectral ones. It is commonly known that in the thermal regime $l_i \ll d$ the
temperature inside PC increases with the bias voltage according to the Kohlrausch relation \cite{PCSbook,Verkin}:

\begin{equation}
\label{eq3}
T_{PC}^{2}=T_{bath}^{2}+V^{2}/4L_{0},
\end{equation}

where T$_{PC}$ is the temperature in the PC core, T$_{bath}$ is the temperature of the bath
and $L_{0}$ is the Lorentz number $(L_{0}=2.45 \cdot 10^{-8\,}$V$^{2}$/K$^{2})$.
In this case, the shape of \textit{I(V)} characteristic for PC is determined by
the bulk resistivity $\rho (T)$ according to Kulik's thermal model \cite{Verkin,Kulik1} as:

\begin{equation}
\label{eq4}
I(V)=Vd\int\limits_0^1 \frac{dx}{\rho (T_{PC} \sqrt{(1-x^2)})}.
\end{equation}

\begin{figure}[t]
\includegraphics[width=8cm,angle=0]{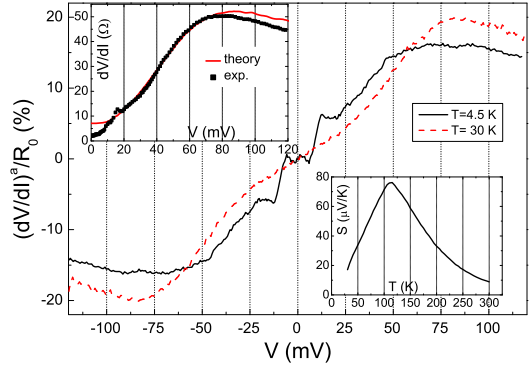}
\caption{(Color online) Calculated antisymmetric part
$dV/dI(V)^{a} =100$\textit{(R(V \textgreater }0)$-$\textit{R(V \textless }0))$/$2$R(V=0)$
of PC from Fig.\,\ref{fig6} with $R$=40\,$\Omega$ at two different temperature 4.5 and 30\,K.
Right inset: Temperature dependence of the thermopower $S$ in Ba$_{0.7}$K$_{0.3}$Fe$_2$As$_2$
\cite{Yan}. Left inset: Symmetrized \textit{dV/dI(V)} for the PC with $R$=1.8$\Omega$ (points)
from Fig.\,\ref{fig6} measured at $T$=4.5\,K along with the calculated \textit{dV/dI(V)} according to Eq.\,(\ref{eq2})
(solid line). To fit the position of the maxima we used an enhanced Lorenz number $L$=2.8$L_{0}$ in the calculation. }
\label{fig7}
\end{figure}

First, we have calculated \textit{dV/dI(V)} according to Eq.\,(4) for several PCs
in the thermal regime within this model using the temperature dependence of
the bulk resistivity $\rho$(T) from \cite{Aswartham} for the investigated system.
The obtained result for one of PC (left inset in Fig.\,\ref{fig7})
demonstrates the good qualitative and quantitative correlation with experimental data in the bias region above the
SC peculiarities. This proves the realization of the thermal regime. For this calculation we used the following parameters:
the residual resistivity $\rho _{0}^{PC} \approx$ 80$\mu \Omega$cm, $d$=140\,nm, $L$=2.8$L_{0}$.
Note, that the Lorentz number $L$ is larger than the standard value.  An enhanced $L$ may be due
to the additional contribution of phonons to the thermal conductivity of PC through electrically non conductive osculant surfaces.

We should note that measured \textit{dV/dI(V)} characteristics of the investigated compound are
asymmetric having larger \textit{dV/dI(V)} values for the positive bias. The similar asymmetry was
also reported in \cite{NaidyukBanafeas} for the samples with $x$=0.25.  In Fig.\,\ref{fig7},
we present the antisymmetric part of $dV/dI(V)^{a} \equiv R^{a}(V)$
for the PC from Fig.\,\ref{fig6}. The calculated $R^{a}(V)$ demonstrates a broad maximum
at about 70--80\,mV.
Qualitatively, the shape of the latter corresponds well to the temperature dependence of
the thermopower \textit{S(T)} measured for the isostructural compound Ba$_{0.7}$K$_{0.3}$Fe$_{2}$As$_{2}$ \cite{Yan}
(see right inset in Fig.\,\ref{fig7}). This fact points out to that the PCs at high bias
are certainly in the thermal regime while at the low bias the spectral regime
is possible. Detail analysis of the thermal regime is given in \cite{NaidyukBanafeas} for
the sample with $x$=0.25.

\begin{figure}[t]
\includegraphics[width=8.5cm,angle=0]{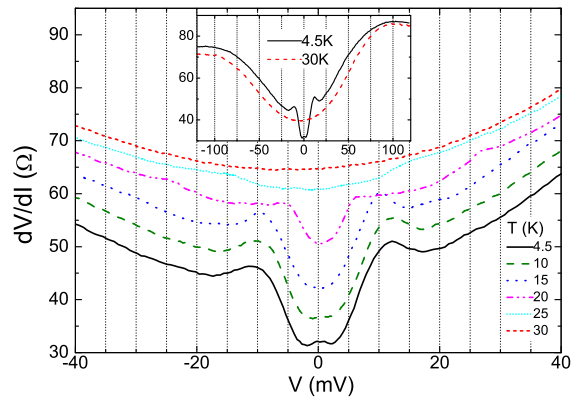}
\caption{ (Color online) $dV/dI(V)$ curves measured for the Ba$_{0.65}$Na$_{0.35}$Fe$_2$As$_2$--Cu
PC from Fig.\,\ref{fig6} with $R \approx  40\,\Omega$ at various temperatures. The curves, except the bottom one,
are shifted for better vizualization.
Inset: \textit{dV/dI} of this PC recorded in a wider bias range for the lowest (4.5\,K) and highest
(30\,K) temperature.}
\label{fig8}
\end{figure}
Let's turn to $dV/dI(V)$ curves with AR-like shape shown on Fig.\,\ref{fig6}. Figure\,\ref{fig8} presents the temperature
measurements of the \textit{dV/dI(V)} in a wide temperature range. Two curves measured
at lowest and highest temperature in a wide voltage range are shown separately in the inset. Like in the case of compounds with
$x$=0.25 \cite{NaidyukBanafeas}, the \textit{dV/dI(V)} curves possess a pronounced asymmetry and display
high-bias maxima (here at about $\pm$100\,mV). Besides, at the lowest temperatures \textit{dV/dI(V)}
displays zero-bias minima, which are likely due to Andreev reflection. At the temperature rise they
transform to a single minimum with the decrease of its amplitude. Finally it vanishes above 25\,K,
what is close to $T_{c}$ of the bulk sample for $x$=0.35. Calculation of the SC gap using conventional
BTK fit procedure \cite{Daghero2011} results in $\Delta\simeq$3.7\,meV for this PC and the reduced
gap value 2$\Delta/$k$_B$T$_c \simeq 2.8-3.4$ if we take T$_c$ in the range 25--30\,K, where the dip minimum
in $dV/dI$ on in Fig.\,\ref{fig8} disappears.
Remind, that above we paid attention that this PC may be affected by heating with voltage increase
and the developed peaks around $\pm 10\,$mV may influence the gap determination.

\begin{figure}[t]
\includegraphics[width=8cm,angle=0]{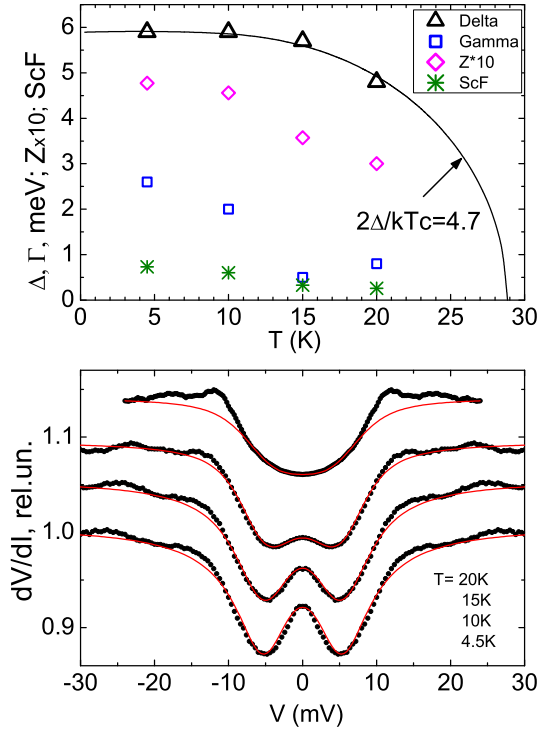}
\caption{ (Color online). Upper panel: Temperature dependence of the SC gaps $\Delta$,
broadening parameters $\Gamma$, barrier strengths $Z$ and scaling factor (ScF) (see Appendix in
Ref.\,\cite{Naid11} to learn the meaning of these parameters)  received from the BTK fit
for the PC ($R\simeq75\,\Omega$) from Fig.\,\ref{fig6}.
Solid line is BCS-like curve.
Bottom panel:
Examples of the symmetrized experimental \textit{dV/dI} (points) and calculated ones (lines) within
the BTK model  \cite{Daghero2011} for several temperatures. } 
\label{fig9}
\end{figure}

So, we have carried out similar procedure (fit) for a PC ($R_N\simeq 75\,\Omega$) from Fig.\,5 with more
pronounced AR minima.
The results are shown in Fig.\,8. For this PC $\Delta\simeq$6\,meV is larger
and the reduced gap value 2$\Delta/$k$_B$T$_c \simeq 4.7$, where T$_c$ is taken from the BCS
curve in  Fig.\,8, because this PC has not survived the temperature increase above 20\,K.
The reduced gap value estimated from the BTK fit using the local T$_{c}$ in other PCs results
in average to 2$\Delta/$k$_B$T$_c$ = 3.6$\pm$1.
As we mentioned above for the compound with $x$=0.25 the larger value 2$\Delta/$k$_{B}T_{c}\approx$6
was found. So, the difference between these two compounds is that for sample with $x$=0.25
superconductivity coexist with spin-density wave order.

The obtained coupling strength values correspond well to those 2$\Delta/$k$_B$T$_c$ = 2.5--4
for another hole-doped system from this family Ba$_{0.55}$K$_{0.45}$Fe$_2$As$_2$ measured
by the same PCAR technique in \cite{Szabo}.
Besides, the lower value of 2$\Delta/$k$_B$T$_c$ = 2.0--2.6 is extracted for the
similar system Ba$_{0.6}$K$_{0.4}$Fe$_2$As$_2$ by PC study in \cite{XinLu}.
At the same time, the gap value obtained from ARPES measurements \cite{Aswartham} for the
compound with the highest T$_c$ ($x$=0.4) is maximal (around 10.5\,meV) for the
inner $\Gamma$ barrel and minimal (around 3\,meV) for the outer $\Gamma$ barrel,
However, as in the case of sample with $x$=0.25, we could not resolve unequivocally the second gap
features. Probably, due to short mean free path of electrons caused by strong elastic scattering
we measure some averaged gap by PCs or, it is not excluded, merging of small and large gaps.
Also scanning tunneling spectroscopy measurements \cite{Song} carried out on similar
compound Sr$_{0.75}$K$_{0.25}$Fe$_2$As$_2$ resolved only one gap, which variates by 16\%
on a 3\,nm length scale, with average 2$\Delta/$k$_B$T$_c$ = 3.6.


\section{CONCLUSIONS}
PC studies of Josephson effect and Andreev reflection were carried out on the iron-based
superconductor Ba$_{1-x}$Na$_{x}$Fe$_{2}$As$_{2}$ with $x$=0.35 and 0.25.
We succeed to detect and study the $ac$ Josephson effect in PCs between Ba$_{1-x}$Na$_{x}$Fe$_{2}$As$_{2}$
crystals and counterelectrodes from Pb, Ta and Nb. Correspondence of the distance between
Shapiro steps to the relation 2$eV=\hbar\omega$ and observed critical current dependence
on the microwave power (ScS with of Pb) support the presence of the s-wave singlet
pairing symmetry in this new superconductor.

Analysis of the measured PC $dV/dI$ spectra with Andreev-reflection features
shows that at small biases the diffusive regime of  the current flow realizes,
whereas at the further bias increase the transition to the thermal regime in PCs occurs.
In the latter case the shape of \textit{dV/dI} is defined by the specific
resistivity $\rho(T)$ of the investigated sample and the noticed
asymmetry of the PC spectra is influenced by the thermopower \textit{S(T)} behavior.
In the spectroscopic regime far below $T_c$ it was possible to detect AR-like
zero-bias minima in the range $\pm$(5--15)\,mV. Applying of the BTK fit gives
the reasonable values of 2$\Delta/$k$_B$T$_c$ = 3.6$\pm$1.
The features on $dV/dI$ which could be due to the manifestation of the second
SC gap appeared very seldom and were not reproducible. This do not allow us
to make a detailed numeric analysis of the second SC gap.
 	
\section*{Acknowledgements}
Funding by the National Academy of Sciences of Ukraine under project $\Phi $3-19
is gratefully acknowledged. Yu.G.N. and O.E.K. would like to thank the IFW Dresden
for hospitality and the Alexander von Humboldt Foundation for the financial support
in the frame of a research group linkage program. S.W. acknowledges the Deutsche
Forschungsgemeinschaft DFG (priority program SPP 1485 and Emmy Nother program;
projects BU887/15-1 and WU595/3-2) for support.  We thank B. B\"uchner for valuable discussions.

\end{document}